\documentstyle[12pt]{article}
\def\beq{\begin{eqnarray}}
\def\eeq{\end{eqnarray}}
\def\ii{\'{\i}}
\def\nl{\newline}
\def\ni{\noindent}
\def\Dc{{\cal D}}
\def\no{\nonumber}
\def\be{\begin{equation}}
\def\ee{\end{equation}}
\begin{document}
\begin{center}
{\Large {\bf Triplectic Gauge Fixing for N=1 Super Yang-Mills Theory}}\\
\vspace{1cm}
C. N. Ferreira \footnote{crisnfer@cbpf.br} and C. F. L. Godinho \footnote{godinho@cbpf.br}  \\
\vspace{1cm}

{\it
 Centro Brasileiro de Pesquisas F\ii sicas (CBPF) -CCP,\\
Rua Dr. Xavier Sigaud 150 - Urca, 22.290-180, Rio de Janeiro,
RJ, Brazil}

\vspace{1cm}

\end{center}

\vspace{1cm}

\abstract
The $Sp(2)$-gauge fixing of $N = 1$ super-Yang-Mills theory is considered here.
We thereby apply the triplectic scheme, where two classes of gauge-fixing bosons are
introduced.  The first one  depends only on the gauge field, whereas the second
boson depends on this gauge field and also on a pair of Majorana fermions.
In this sense, we build up the BRST extended (BRST plus antiBRST) algebras for
the model, for which the nilpotency relations,
$s^2_1=s^2_2=s_1s_2\,+\,s_2s_1=0$, hold.

\vskip3cm
\noindent PACS: 11.15 , 03.70

\vfill\eject
\section{Introduction}
In globally  supersymmetric  gauge theories,  gauge choices may be adopted that break
supersymmetry; that is the case, for example, of the Wess- Zumino gauge choice.
This sort of breaking is not spontaneous.  Indeed, supersymmetry 
becomes nonmanifest and, in some cases, 
it may jeopardize the quantization program, for nonphysical
states may appear in the spectrum. Such an issue may be very 
systematically treated by means of the Becchi-Rouet-Stora-Tyutin- (BRST-) extended quantization
procedure by Batalin and Marnelius, also referred to as the triplectic
scheme \cite{T1,T2,T3}.

The triplectic scheme is a
general covariant $Sp(2)$-symmetric Lagrangian gauge theory
quantization procedure that follows
the general procedure of the field-antifield or
Batalin Vilkovisky (BV) method\cite{BV1,BV2};
however, it relies on with the additional requirement of an extended
BRST\cite{He1,He2,He3} (BRST plus anti-BRST) invariance, rather
than just BRST symmetry.
In the usual BV quantization, the BRST invariance is translated
into the so-called master equation. At zero-loop order, this
equation is well-defined and its solution, together with the
appropriate requirements corresponding to the gauge fixing,
leads to the construction of the complete structure of ghosts,
antighosts, ghosts for ghosts, etc\cite{HT,H1}.
At higher orders in $\hbar$, one needs however to introduce some
regularization procedure in order to give  a well-defined meaning
to the mathematical objects involved in the formal master
equation. Anomalies and Wess Zumino terms may in this way  be
calculated at one-loop order\cite{TPN,DJ}.

In the triplectic quantization, the extended BRST invariance
is expressed through a set of two master equations
that correspond to the requirements of BRST and anti-BRST
invariances,
respectively. As in the standard BV case, both
equations formally admit a loop expansion.
One then expects that anomalies and Wess Zumino terms should  show up
at one-loop order, as long as one is able to introduce appropriate
regularization schemes.
These features are not manifest in the recently discussed case of the
Yang Mills theory presented in ref\cite{ABG}.

Our proposal in this paper is to discuss the gauge fixing of $N=1$\,- SYM model by
consider\-ing the background-field procedure.  Firstly, we obtain the complete
structure for the BRST extended symmetry (BRST plus anti-BRST).
This algebraic structure will come to help us in the
development of two classes of gauge-fixing bosons.  The first
class boson has a dependence only on the gauge field, $A_\mu$, but the second class
boson is a function of $A_\mu$ and the  $\lambda$ and $ \bar\lambda$ Majorana fermions.
In both cases, we obtain the gauge-fixed action for the model.
We shall also show how to fix the gauge by means of canonical transformations
by considering the method of ref. \cite{ABG}.

\section{Review}
The field-antifield formalism for quantization of general dynamical systems is
the most powerful method  to treat gauge models. In this way, we start
by considering some gauge theory, and enlarge the original field
content,
$\,\phi^i\,$, adding up all the usual gauge-fixing structure: ghosts,
antighosts and auxiliary fields associated to the original gauge
symmetries. The resulting set will be labeled by $\,\phi^A\,$.
Then, we associate to each of these fields five new quantities,
introducing the sets:  $ {\bar \phi}^A $, $\,\phi_A^{\ast\,1}\,$ ,
$\,\phi_A^{\ast\,2}\,$,$\,\pi_A^{1}\,$ and  $\pi_A^{2}\,$.
The Grassmannian parities of these fields are:
$\epsilon (\phi^A) \,=\,\epsilon ({\bar \phi}^A)
\,\equiv \,\epsilon_A\,$,
$\epsilon (\phi^{\ast\,a}_A ) \,=\,\epsilon ( \pi^a_A ) \,=\,
\epsilon_A\,+\,1\,$.
The ideas of the extended BRST quantization in the antifield
context previously discussed in\cite{BLT1,BLT2,BLT3} are brought into a
completely anticanonical setting.
The extended BRST invariance of the generating
functional, defined on this $6n$-dimensional space, is equivalent to
the fact that the quantum action, $W$, is a
solution of the two master equations:

\begin{equation}
\label{ME}
{1\over 2}\big( W \,,\, W\,\big)^a \,+\,V^a W \,=\,
i\hbar \Delta^a W,
\end{equation}

\noindent where the indices $\,a\,=\,1\,,\,2\,$ correspond
respectively
to BRST and anti-BRST invariances and the extended form of the
antibrackets, triangle and $V$ operators read as below:

\begin{equation}
\label{AB}
\big( F\,,\,G\,\big)^a\,\equiv \,{\delta^r F\over \delta \phi^A}
{\delta^l G \over \delta \phi_A^{\ast\,a}}\,+\,
{\delta^r F\over \delta {\bar \phi}^A}
{\delta^l G \over \delta \pi_A^a\,}
\,-\, {\delta^r F \over \delta \phi_A^{\ast\,a}}\,
{\delta^l G\over \delta \phi^A}\,-\,
{\delta^r F \over \delta \pi_A^a\,}
{\delta^l G\over \delta {\bar \phi}^A}
\end{equation}

\begin{equation}
\Delta^a\,\equiv\, (-1)^{\epsilon_A}\,
 \,{\delta^l \over \delta \phi^A}
{\delta^l  \over \delta \phi_A^{\ast\,a}}\,+\,
(-1)^{\epsilon_A}\,
{\delta^l \over \delta {\bar \phi}^A}
{\delta^l  \over \delta \pi_A^a\,}
\end{equation}

\begin{equation}
V^a\,=\,{1\over 2} \epsilon^{ab}\,\Big( \phi_{A b}^{\ast}
{\delta^r \over \delta {\bar \phi}^A}
- (-1)^{\epsilon_A} \pi_{A\,b} {\delta^r \over
\delta \phi^A} \Big)\,\,.
\end{equation}

\noindent Here and in the sequel,
unless explicitly
indicated, we shall be adopting the convention of summing up
over repeated indices.

The field-antifield functional integral is defined including also an extra
action functional, $X$, with a gauge-fixing status:
\begin{equation}
\label{VAC}
Z\,=\, \int [{\cal D}\phi][\Dc \phi^{\ast}][\Dc \pi]
[\Dc \bar\phi ][\Dc \lambda]\,exp\{{i\over \hbar}\big(
W\,+\,X\big)\};
\end{equation}
\ni
this functional is required to satisfy the following master equation
\begin{equation}
\label{ME2}
{1\over 2} \big(X \,,\, X\,\big)^a \,-\,V^a X \,=\,
i\hbar \Delta^a X.
\end{equation}

Another way of gauge fixing is by means of the canonical transformations
rather than including the functional $X$, was proposed in \cite{ABG}.
In
this work, we  subsequently apply this method for the gauge fixing
of an ($N\,=\,1$) SYM theory.

For a gauge theory with closed and irreducible algebra, corresponding
to a classical action $S_0[\phi^i]$, a solution for
the zero-loop order action, $S$, is:

\begin{equation}
\label{TAC}
S \,=\, S_0 + \phi_A^{\ast\,a}\,s_a \phi^A
+ {1\over 2} {\bar \phi}_A s_2
s_1 \phi^A
\,+\,{1\over 2} \epsilon^{ab} \phi_{A\,a}^{\ast}\,\pi^A_b,
\end{equation}

\ni where the $s_a$ represents
BRST ($a=1$) and anti-BRST ($a=2$) transformations of the fields
(in other words, for theories with closed algebra, the standard BRST
extended algebra associated with the gauge theory). In this article, we
shall  not be dealing with the generalized BRST transformations of the
triplectic formalism\cite{T1,T2,T3}, but just with standard transformations
that do not involve the antifields.

\section{Extended BRST Symmetry for N\,=\,1\,-\, SYM}
Traditionally, Lagrangians invariant under supersymmetry and a local gauge symmetry
 also exhibit spin-$1\over2$ fermions and scalar fields.  Here, we are interested in a
 Lagrangian invariant under the smallest degree of supersymmetry, namely $N\,=\,1$, whose
 multiplet displays the gauge boson ($A_{\mu}$) and its physical Majorana fermion
 partner ($\lambda$), the gaugino.  Let us consider the classical action for this
 model:

\be
S_{0}\,=\,\ - \frac{1}{g^2} 2tr\int d^4x\big({1\over4}F_{\mu\nu}F^{\mu\nu}\,+
\,i\bar{\lambda}\bar{\sigma^{\mu}D_{\mu}\lambda}  \big),
\ee
where $g$ is the gauge coupling, and we adopt:

\begin{eqnarray}
D_{\mu}\bullet\,=\,\partial_{\mu}\bullet\,-\,i[A_{\mu},\,\bullet] \no\\
F_{\mu\nu}\,=\,\partial_{\mu}A_{\nu}\,-\,\partial_{\nu}A_{\mu}\,-\,i[A_{\mu}\,,\,A_{\nu}].
\end{eqnarray}

\noindent
The $N\,=\,1$ SUSY algebra reads as below:
\label{SUSY}
\beq
\delta^{SUSY} A_{\mu}\,=\, i(\bar{\xi}\bar{\sigma}_{\mu}\lambda\,-\,\bar{\lambda}\bar{\sigma}_{\mu}\xi) \no\\
\delta^{SUSY} \lambda\,=\,\sigma^{\mu\nu}\xi F_{\mu\nu}\no\\
\delta^{SUSY} \bar{\lambda}\,=\,-\bar{\xi}\bar{\sigma}^{\mu\nu}F_{\mu\nu}
\eeq

An appropriate form to write down the BRST transformations may be
found in \cite{FUJ}, and a BRST extended (BRST $s_1$ plus anti-BRST $s_2$) version satisfying the extended nilpotency,
\nl
\be
s_1^2\,=\,s_2^2\,=\,s_1s_2\,+\,s_2s_1\,=\,0,
\ee
may be written as
\beq
\label{extended}
s_1A_{\mu}\,=\,i D_{\mu}c_1 \no\\
s_1\lambda\,=\,-[c_1,\,\lambda] \no\\
s_1\bar{\lambda}\,=\,-[c_1\,,\,\bar{\lambda}]\no\\
s_1c_1\,=\, -{1\over2}[c_1\,,\,c_1] \no\\
s_1c_2\,=\,-B \no\\
s_1B\,=\,0;
\no\\
\no\\
s_2A_{\mu}\,=\,iD_{\mu}c_2 \no\\
s_2\lambda\,=\,-[c_2\,,\,\lambda]\no\\
s_2\bar{\lambda}\,=\,-[c_2\,,\,\bar{\lambda}] \no\\
s_2c_1\,=\,B\,-\,[c_1\,,\,c_2]\no\\
s_2c_2\,=\,-{1\over2}[c_2\,,\,c_2] \no\\
s_2B\,=\,[B\,,\,c_2];
\eeq
we take here $c_1\,=\,c_1^aT^a\,,\,c_2\,=\,c_2^aT^a$ and $B\,=\,B^aT^a$ for ghosts,
antighosts and auxiliary fields.  Also, we adopt the normalization condition
$Tr(T^aT^b)\,=\,{1\over2}\delta^{ab}$.

An important point to consider now is the general form of the gauge-fixing
action in the triplectic scheme \cite{T1}.  The method consists in the construction
of a non-degenerate  (gauge-fixed) action that belongs to the same cohomological class
as the classical action;  consequenctly, it describes the same physical observable \cite{HT}.
The gauge fixing with the BRST extended invariance has been analyzed in many papers
\cite{He1,He3,ABG,W2,BLT1,BLT3}.  A most general form of the BRST extended gauge-fixed
action has been analyzed in Hamiltonian framework, in \cite{He1}.  In this paper,
we claim that the gauge-fixed action does not have the general  form
 $S_{GF}\,=\,S_0\,+\,s_2s_1\chi$.  We will however be interested just in the
 $Sp(2)$-symmetric case described in Section 2, for which this results hold.

Let us consider the triplectic functional of (\ref{VAC}) with a suitable
gauge-fixing functional, $X$, solution to the eq.(\ref{ME2}).  After integrating
over $\bar\phi_A\,,\phi^{\ast}_{Aa}$, $\pi^a_A$, the conclusive result will be the
exponential of ${i}\over{\hbar}$ times a (non-degenerated) gauge-fixed action
of the form,

\be
\label{AGF}
S_{GF}\,=\,S_0[\phi^i]\,+\,s_2s_1\chi[\Phi^A].
\ee

This action is trivially BRST-extended invariant.  The precise relation between the
bosonic functional, $\chi$, and the triplectic gauge fixing action, $X$, of (\ref{ME2})
is not relevant for our purposes here.  The relevant question in our study is to obtain the
gauge-fixed action for the $N\,=\,1$ supersymmetric Yang-Mills model by using this framework
of the background.  In this way, some questions may appear:  How to construct the gauge
fixed-boson?  What is its functional dependence?
\nl
The heart of our analysis is that, for the gauge-fixing action (\ref{AGF}), the SUSY
charge density, $J^0$, is
\be
J^0\,=\,J^0_{Naive}\,+\,BRST\, extended\,\, exact,
\ee
if
\be
\delta^{SUSY} s_2s_1\chi\,=\,s_2s_1\delta\chi ,
\ee
where $\delta$ is a supersymmetry transformation and $s_2s_1$ is the BRST-extended piece.
A recent paper, \cite{FUJ}, shows how to gauge-fix this model in the BRST framework; for
this purpose, the authors use the well-known fermion, $\Psi\,=\,c_2\partial_{\mu}A^{\mu}$.
\newline
Our purpose is to obtain a bosonic functional of $A_{\mu}\,,\lambda\,\,\, and\,\,\, \bar{\lambda}$
fields.
We can easily show that SUSY and BRST extended symmetries commute with these
quantities, by considering the algebras (\ref{SUSY}) and ({\ref{extended}), where
\beq
\delta^{SUSY} s_1s_2 A_{\mu}\,=\,i[B\,,\,\bar{\xi}\bar{\sigma_{\mu}}\lambda\,-\,\bar{\lambda}\bar{\sigma_{\mu}}\xi]\,+\,[[\bar{\xi}\bar{\sigma_{\mu}}\lambda\,-\,\bar{\lambda}\bar{\sigma_{\mu}}\xi\,,c_{1}]\,,\,c_{2}] \no \\
\delta^{SUSY} s_1s_2 \lambda\,=\,-\sigma^{\mu\nu}\xi[F_{\mu\nu}\,,\,B]\,+\,\sigma^{\mu\nu}\xi[F_{\mu\nu}\,,\,c_1]c_2\,+\,\sigma^{\mu\nu}c_2[F_{\mu\nu},c_1] \no \\
\delta^{SUSY} s_1s_2 \bar{\lambda}\,=\,\bar{\xi}\bar{\sigma_{\mu\nu}}[F_{\mu\nu}\,,\,B]\,-\,\bar{\xi}\bar{\sigma_{\mu\nu}}c_2[F_{\mu\nu}\,,\,c_1]\,-\,\bar{\xi}\bar{\sigma_{\mu\nu}}[F_{\mu\nu}\,,\,c_1]c_2 .
\eeq
These identities allow us to write a gauge-fixing boson in two different cases.
\newline
(I) Let us start the first and particular case for the boson that involves only the field
$A_{\mu}$:
\be
\label{BOS1}
\chi\,=\,-{1\over2}A_{\mu}A^{\mu}
\ee
(II) In the second and most general case, the boson is a functional of the $A_{\mu},\lambda,\,\,and\,\,\,\bar{\lambda}$
fields:
\be
\label{BOS2}
\chi\,=\,-{1\over2}A_{\mu}A^{\mu}-{1\over2}\lambda \bar{\lambda}
\ee
In both cases, we find the usual gauge-fixed action for this
model:
\be
s_1s_2 \chi\,=\,tr(\partial_{\mu} B A^{\mu}\,+\,i\partial_{\mu} c_2 D^{\mu}c_1).
\ee

Another interesting way to get the gauge-fixed action is to perform canonical
transformations in the triplectic space \cite{ABG}.  For each of the antibrackets
of (\ref{AB}), we introduce a generator of transformations
$F_a\,[\phi^A\,,{\bar \phi}^A\,,\phi^{\ast\,a\,\prime}_A\,,
\pi^{a\,\prime}_A\,\,]\,$ and write down the set of
transformations:
\begin{eqnarray}
\label{CTR1}
\phi^{A\,\prime} &=& {\delta F_a \over \delta \phi^{\ast\,a\,\prime}_A}
\no\\
\phi_A^{\ast\,a} &=& {\delta F_a \over \delta \phi^A}\no\\
{\bar \phi}^{A\,\prime} &=& {\delta F_a \over \delta \pi^{a\,\prime}_A}
\no\\
\pi_A^{a} &=& {\delta F_a \over \delta {\bar \phi}^A}\,,
\end{eqnarray}

\noindent
where their general form is
\begin{equation}
F_a \,=\,{\bf 1}_a\,+\,f_a
\end{equation}

\ni with

\begin{eqnarray}
{\bf 1}_a &=&  \phi^A \phi^{\ast\,\prime}_{A\,a}
\,+\,{\bar \phi}^A \pi^{\prime}_{A\,a} \no\\& &\no\\
f_1 &=& g_1 [\phi\,,{\bar \phi}]\,+
\,g_3^A[\phi\,,{\bar \phi}] \pi^{ 1\,\prime}_A \,+\,
g_4^A [\phi\,,{\bar \phi}]\phi^{\ast\,1\,\prime}_A \no\\& &\no\\
f_2 &=&  g_2[\phi\,,{\bar \phi}]
\,+\, g_3^A [\phi\,,{\bar \phi}] \pi^{2\,\prime}_A\,+\,
g_4^A [\phi\,,{\bar \phi}]  \phi^{\ast\,2\,\prime}_A \,\,.
\end{eqnarray}

In this approach, the fundamental condition for the canonical transformation reproduces
the gauge-fixing corresponding to some boson $\chi$, after we express the result in terms
of the transformed fields and impose the condition that
$\bar{\phi_A}^{\prime}\,,\,\phi^{\star \prime}_A$ and $\pi^{a \prime}_A$
are set to zero:
\begin{equation}
\label{general}
{\delta f^{\prime}_a\over \delta \phi^A }s_a
\phi^A \,+\,{1\over 2} g^{\prime}_3 s_2 \,
s_1 \phi^A \,-\,
{1\over 2}\epsilon^{ab} {\delta f^{\prime}_a\over \delta \phi^A }
{\delta f^{\prime}_b\over \delta {\bar \phi}^A }\,\,=\,
 s_2 s_1 \chi \,\, .
\end{equation}

For the present case, we may to consider two possibilities
\newline
(I)
\newline
\begin{eqnarray}
g_1 &=& g_3 \,=\, g_4 \,=\, 0 \no \\
g_2 &=& s_1 \chi_i \no \\
f_2 &=& s_1\chi_i
\end{eqnarray}

\noindent or
\newline
(II)
\newline
\begin{eqnarray}
g_2 &=& g_3 \,=\, g_4 \,= \, 0 \no \\
g_1 &=& -\,s_2 \chi_i \no \\
f_1 &=& -\,s_2 \chi_i ,
\end{eqnarray}
where $i\,=\,1,2$ refers to the two classes of bosons (\ref{BOS1}) and (\ref{BOS2}).
In both cases, we get the gauge-fixed action, whenever we perform the corresponding
canonical transformation in the fields and then set all the primed antifields to zero.

\section{Concluding Remarks}
We have shown that it is possible to obtain the gauge-fixed action for an $N=1\,-$ Supersymmetric
Yang-Mills model with BRST extended invariance.  To carry out such a programme, we consider two paths for the
gauge-fixing.  In the first one, we adopt the triplectic scheme where, for a particular solution
to the gauge-fixing action $X$, two kinds of bosons are considered; in both cases, we obtain the
correct form of the gauge-fixed action.  In the second way, we perform the gauge-fixing process
with the help of canonical transformations by means of a particular choice for the
generator of the transformations.
\vspace{2cm}

\section{Acknowledgments} The authors would like to express their deepest gratitude to
 J. A. Hela\"yel-Neto for a careful reading and suggestions on
our original manuscript. CNPq-Brazil is acknowledged for the
invaluable financial help.

\vfill\eject

\end{document}